UNIVERSITY OF
BIRMINGHAM

# COMMAND & CONTROL
## UNDERSTANDING, DENYING AND DETECTING

JOSEPH GARDINER
MARCO COVA
SHISHIR NAGARAJA

FEBRUARY 2014

# Abstract

One of the leading problems in cyber security today is the emergence of targeted attacks conducted by adversaries with access to sophisticated tools, sometimes referred to as Advanced Persistent Threats (APTs). These attacks target specific organisations or individuals and aim at establishing a continuous and undetected presence in the targeted infrastructure. The goal of these attacks is often espionage: stealing valuable intellectual property and confidential documents.

As trends and anecdotal evidence show, providing effective defences against targeted attacks is a challenging task. In this report, we restrict our attention to a specific part of this problem: specifically, we look at the Command and Control (C2) channel establishment, which, as we will see, is an essential step of current attacks. Our goals are to understand C2 establishment techniques, and to review approaches for the detection and disruption of C2 channels.

More precisely, we first briefly review the current state of cyber attacks, highlighting significant recent changes in how and why such attacks are performed. This knowledge is foundational to understand C2 techniques and to design effective countermeasures.

We then investigate the "mechanics" of C2 establishment: we provide a comprehensive review of the techniques used by attackers to set up such a channel and to hide its presence from the attacked parties and the security tools they use.

Finally, we switch to the defensive side of the problem, and review approaches that have been proposed for the detection and disruption of C2 channels. We also map such techniques to widely-adopted security controls, emphasizing gaps or limitations (and success stories) in current best practices.

**CPNI**
Centre for the Protection
of National Infrastructure

We would like to acknowledge the help and support of CPNI in researching this topic and producing the accompanying products.

# Executive Summary

Formation and use of a Command and Control (C&C) system is an essential part of remotely-conducted cyber attacks. C&C is used to instruct compromised machines to perform malicious activity - C&C can also be used as a channel over which data can be exfiltrated. Statistics show that cyber attacks are widespread across all sectors and that preventing intrusion is difficult. A promising alternative consists of detecting and disrupting the C&C channels used by attackers: this effectively limits the damage suffered as a consequence of a successful attack (e.g., preventing sensitive data to being leaked).

## C&C communication and traffic

Attackers experiment with alternative strategies to build reliable and robust C&C infrastructures and to devise stealthy communication methods. As a consequence, different C&C architectures and communication techniques have emerged. For example, attackers have used centralised architectures, based on the standard IRC and HTTP protocols. More recently, they have introduced decentralised architectures based on P2P protocols, which are more difficult to take down. Similarly, direct forms of communication have been substituted by encrypted channels, where attacker's commands and stolen information cannot be readily accessed. To make channel detection and blocking more difficult, attackers also use covert communication mechanisms that mimic regular traffic patterns. For example C&C traffic can occur through pages and images on Online Social Networks (OSNs), covert DNS traffic, and networks for anonymous communication, such as Tor.

## C&C detection and disruption

A variety of techniques for the detection and disruption of C&C channels have been proposed. They typically rely on the automated monitoring and analysis of network traffic to identify indicators of compromise, malicious traffic, or anomalous communication patterns. The importance of human involvement in this activity cannot be overstated. As attackers constantly adapt their strategies, it is critical to gain a thorough understanding of the traffic flow patterns followed by manual tuning of monitoring, detection, and response infrastructure at periodic intervals.

The following is a checklist of measures that help detecting and denying C&C in your organisation.

### Detect known-bad network activity

Collect and analyse network traffic to identify activity that is known to be caused by an active C2 channel.

- *Monitor DNS traffic* to identify internal devices that attempt to contact domains that are known to be involved in C2 activity. This measure involves collection of DNS traffic information (either through a passive DNS collector or via the nameservers logs) and matching of requests against one or more blacklists of malicious domain names.

- *Monitor IP traffic* to identify internal devices that attempt to connect to end points that are known to be involved in C2 activity. This measure involves collection of IP traffic information (for example, enabling NetFlow and sFlow collection in routers) and matching of communications against one or more blacklists of malicious IP addresses.

- *Monitor traffic content* to identify content that matches known C2 traffic (e.g., specific network request/responses signatures). This measure involves collection of full traffic content (for example, enabling a network sniffer) and matching of the collected data against traffic signatures.

These measures enable the detection of C2 channels that are set up by known malware families, leverage known infrastructure, or employ known communication techniques.

### Detect anomalous network activity

Collect and analyse network traffic to identify activity that deviates from the expected, normal traffic profile of the monitored network.

- *Establish traffic baselines* to determine the "normal" profile of the network (normal communication patterns, data exchange volumes, etc.). This measure can be implemented by determining baselines for different time windows (e.g., hour, day), internal devices, and network services.

- *Evaluate current network activity* against the established baselines to identify deviations that may be indicative of C2 activity. Pay particular attention to anomalies such as periodic beaconing, surge in the amount of exchanged traffic, suspicious network behaviours.

- *For example*, C2 activity that relies on fast-flux techniques can be detected by searching DNS data for patterns of fast-changing associations between domain names and IP addresses; DGA-based C2 activity is revealed in DNS data by use-and-discard patterns of domain names; data exfiltration may be detected in Net-Flow data by unusually large volumes of data exchanges.

These measures enable the detection of C2 channels that are set up by never-seen-before malware families and that do not re-use any known malicious infrastructure.

### Deny C2 activity

Architect and operate the network in such a way that C2 activity is effectively denied or greatly impaired.

- *Segment the network* to separate devices with different trust and risk values (e.g., front-facing, publicly servers vs. internal hosts storing sensitive documents).

- *Introduce rate-limit policies* to slow down traffic directed to disreputable or un- trusted endpoints.

- *Block unwanted or unused communications mechanisms* that may be used to piggy back C2 activity (e.g., anonymisation networks, P2P overlays, social net-works).

# Executive Summary

## Practicalities

Start small, measure, and scale up: security controls can be applied itertively, covering first high-risks groups, idetifying mechanisms that are effective, and then expanding their applications to larger portions of the organisation.

# Introduction

We are currently in the middle of a computer security crisis: the number of attacks, their sophistication and potential impact have grown substantially in the last few years. In particular, *targeted attacks*, sometimes also called advanced persistent threats (APTs), have emerged as today's most challenging security threat. Targeted attacks target specific individuals or organisations with the typical intent of obtaining confidential data, such as contracts, business plans, and manufacturing designs. They typically employ extensive reconnaissance and information gathering to identify weaknesses in the target's defences, and rely on sophisticated malware to perform the intended actions (e.g., locate and steal sensitive documents within the target's network).

Because of their nature, targeted attacks are particularly difficult to prevent. To in- trude and take control of the target's systems, they may use 0-day exploits [11] or other malicious code that is known to evade the specific defence mechanisms used by the target. They may also rely on carefully-crafted social engineering techniques to "exploit the human", that is to convince unsuspecting users within the targeted organisation to perform unwanted activities, such as installing and running malware.

An additional line of defence against targeted attacks is the detection and disruption of individual steps that are essential for the successful progression of an attacks. This is the so-called kill chain approach [19]. Of particular interest for a defender is identifying the step in which a compromised system establishes a Command & Control channel (C2), i.e., a communication channel with the attackers through which it can receive further commands or can send any stolen data.

Blocking an intrusion in the C2 step has several advantages. If no sensitive data is ever exfiltrated, the targeted organisation limits its damage significantly: while the integrity of the organisation's systems has been compromised, its most valuable assets (e.g., intellectual property and R&D plans) are still intact. Even in the event of successful data stealing, an understanding of the C2 structure could prove essential to determine what has been stolen and where it ended to. In addition, the analysis of the C2 channel may provide indications useful to attribute the attack to specific groups of people, which may facilitate legal actions against them.

The overarching goal of our study is to understand the techniques of Command and Control in order to improve our defensive approaches. We will examine C2 activity both from the attacker's perspective *(how are C2 channels set up and maintained?)* and from the defender's perspective *(how are C2 channels detected and disrupted?)*. Having an understanding of both sides of the problem (attacks and defences) is key to understand what attackers are currently capable of doing (or might do in the future) and what defences may be effective against them.

Our approach to the problem of understanding and combating C2 is based on a com- prehensive review, systematization, and contextualization of the substantive work in this area, done by both the academic and commercial community. For the academic work, we focus our attention on publications appearing in top conferences and journals, such as USENIX Security, ACM CCS, IEEE Security & Privacy, and NDSS. For the indus- try work, we review publications at conferences such as RSA and BlackHat, technical reports, and blog postings authored by the main security vendors. Whenever possible, we emphasise practical considerations extracted from these works, with the hope that they may lead to better defence mechanisms to be deployed. The rest of the report is organised as follows: we start by covering some background material on Command & Control (section C). We then review in detail the techniques that attackers use (or may use) to create and maintain C2 channels (section D). We review approaches that have been proposed to detect C2 channels and disrupt them (section E). Finally, we revisit security controls that are commonly adopted by organisations to spotlight those that are more likely to successfully identify and disrupt C2 activity, and to identify any gaps in the current best practices (section F).

# The Command and Control Problem

Command and Control identifies the step of an attack where the compromised system contacts back the attackers to obtain addition attack instructions and to send them any relevant information that has been collected up to that point. To really understand C2 activity, we need to review a number of aspects that, taken together, characterise today's attacks. In particular, we will examine the factors that shape the current attack landscape (*why targeted attacks have become such a threat?*). We also review the actual way in which the attacks attacks are carried out (*how does a targeted attack work?*) and the reasons why C2 activity is a critical step in these attacks. Then we look at the available data on targeted attacks to quantify them and to learn some lessons specifically on C2 activity, before reviewing notable cases of targeted attacks.

## C.1 Attack Landscape

The security field is co-dependent with an adversary. As the adversary's motivations, drivers, or technical means change, so does the entire security landscape. We posit that changes in cyber attacks that have occurred lately (and that affect our ability to defend against them) are largely the result of several significant changes in the techniques and behaviours of attackers. We focus here on three main thrusts: changes in attackers' motivations, the increased targeting of attacks, and their use of evasive techniques.

**Motivations**

The motivations of attackers have changed substantially, transforming their activity from a reputation economy to a cash economy [37]. Long gone are the days when attacks were performed predominantly by individuals with the intent to display their technical skills and to gain "street credibility". The last ten years have seen instead the rise of criminal groups that use Internet-based attacks to make a financial profit. Criminal groups can be well-organised and technically sophisticated. They can often rely on specialised "contractors" for different parts of the attack: for example, they may include a computer programmer for the development of actual attack code and a "cashier" for the monetization of stolen data. Less advanced groups can rely on the wide availability of commoditised attack tools, such as pre-packaged exploit kits [41] or phishing kits [24], which simplify considerably the steps required to launch relatively sophisticated attacks. Notably, the activity of these groups is sufficiently well-established to give rise to active underground markets, where malicious code, stolen goods, tips and tricks are exchanged or sold [35]. An overview of cyber crime evidence for the UK has been recently published [76]. Traditionally, criminal groups have focused on getting access to financial data, such as credit card numbers and online banking account credentials, which can be easily monetized. This activity has been referred to as "cyber crime", since it replicates traditional criminal activities (such as money stealing and fraud) in the online domain. However, more recently attackers have increasingly targeted sensitive data different than financial, focusing primarily on acquiring intellectual property, such as manufacturing designs, legal contracts, etc. These attacks can often be classified as examples of industrial and commercial espionage.

A significant evolution in this line of changes to attackers' motivation is the rise of State-sponsored attacks. With this term are denoted attacks that, for their scope, objectives, and cost, are likely to be mandated and funded by State-level entities. State-level attacks encompass two typical goals: the systematic and comprehensive espionage of other nations' entire economic sectors with the objective of gaining strategic advantage [13], and the sabotage of critical national infrastructure, such as power plants and transportation control systems. The impact and consequences of these attacks have led some commentators to discuss the possibility of cyber wars [18]. The most well-known example of a State-level attack is Stuxnet, a worm believed to be created by the United States and Israel to sabotage a nuclear facility in Iran [69, 106].

**Targeted Attacks**

A second significant change that is relevant to our study of Command and Control is the increasingly targeted nature of attacks. Cyber crime activity is typically opportunistic: attackers cast a wide net and are happy with any target they can capture. More sophisticated attacks, on the contrary, take aim at very specific organisations or individuals and expend significant resources to compromise them.

This change in the mode of attacks has several important consequences. Attackers do not simply move from one potential victim to another, in search of the system that, being least defended, offers the easiest way in. Instead, attackers focus relentlessly on their selected target.

Second, the methodology of attacks change. In particular, the attack life-cycle includes a reconnaissance phase in which the target's security posture and the defensive tools it uses are carefully examined and analysed to identify possible weaknesses [73]. In addition, a targeted compromise attempts to establish its presence on the victim's systems for as long as possible, so as to reap the benefits of the intrusion over time. Consequently, the life cycle commonly includes phases in which the intruder moves "laterally", i.e., gains access to additional systems, and introduces techniques to main- tain the attackers' presence in the intruded system.

Actual attack artefacts, for example, malware samples or network-based attacks, tend to become unique: they are tailored to a specific target and, thus, are less likely to be reused in other attacks. This is problematic for security tools, which sometimes use the observation of the same suspicious artefact in multiple locations as an indication of maliciousness, and for security companies, which may prioritise the investigation of novel attacks and artefacts based on their prevalence. Security researchers are also less likely to develop signatures to match these rarely-seen artefacts.

# The Command and Control Problem

## Evasions

The last aspect of modern attacks that we want to discuss in detail is their increasing use of evasive techniques. Attackers want to stay under the radar for as long as possible, to avoid being detected or raising alerts. To achieve this, they adopt a number of measures that, as we will see, have a significant, negative impact on the effectiveness of a number of traditional defence mechanisms.

### Evading signatures

Traditional defence systems (such as traditional anti-virus and intrusion detection systems) often rely on signatures to detect attacks or malicious code. A signature characterises a known attack by defining its characteristics. For example, in the context of malware, a signature could be a regular expression that matches the bytes found in a specific malicious file. Unfortunately, a number of obfuscation techniques have been proposed (and are used extensively) to counter signature-based detection. For example, polymorphism is a technique that enables an attacker to mutate an existing malicious binary and create a completely new version from it, which retains its original functionality but is undetected by current signatures [52]. The anti-virus vendor Kaspersky recently reported detecting more than 2 unique malicious samples per second, likely the result of extensive application of polymorphic techniques [64].

### Evading dynamic analysis systems

To overcome the limitations of signature-based analysis of malicious code, researchers use dynamic analysis tools, also called sandboxes [31]. These tools execute a binary in an instrumented environment and classify it as either benign or malicious depending on the observed behaviour.
To thwart automated dynamic analysis, malware authors have developed a number of checks (so-called "red pills") to detect the presence of malware analysis tools and popular sandbox environments. When the malware detects indications that a malware analysis system is present, it typically suppresses the execution of malicious functionality or simply terminates [8]. The way in which the checks are implemented depends on the type of malware analysis system that is targeted. One class of checks inspects the runtime environment to determine whether an analysis tool is present. Often, such checks look for files, registry keys, or processes that are specific to individual analysis tools. A second class of checks exploits characteristics of the execution environment that are different between a real host and a virtualised environment [2, 33, 34, 105] or an emulated system [74, 88, 98] (which are frequently used to implement the analysis sandbox). For these checks, small variations in the semantics of CPU instructions or timing properties are leveraged to determine whether a malware process is run in an emulator or a virtual machine (VM).
As another evasive technique, malware may execute its malicious payload or spe- cific parts of its code only when some "trigger" fires, i.e., only when some specific precondition is satisfied [78]. For example, a malware program may check that certain files or directories exist on a machine and only run parts of its code when they do. Other triggers require that a connection to the Internet be established or that a specific mutex object not exist. Other malware becomes active only in a specific date range, when run by a user with a hard-coded username, or if the system has been assigned a precise IP address. Furthermore, some malware listens for certain commands that must be sent over a control channel before an activity is started.
In the next step of the arms race, malware authors have started to introduce stalling code into their malicious programs [66]. Stalling code is executed before any malicious behaviour, regardless of the execution environment. The purpose of such evasive code is to delay the execution of malicious activity long enough so that the automated analysis system stops the analysis having observed benign activity only, thus incorrectly concluding that the program is non-functional or does not execute any action of interest. Of course, on a regular system, the malware would perform all of its malicious behaviour, right after the delay. Stalling affects all analysis systems (virtualised, emulated, or physical machines), even those that are fully transparent. This technique simply leverages the fact that, to analyse a large volume of programs, an analysis system must bound the time it spends executing a single sample to a limited time (in the order of few minutes). To make things worse, malware authors can often craft their programs so that their execution in a monitoring environment is much slower than in a regular system (by a factor of 100 or even more).

### Evading reputation systems

Another defensive approach that has gained traction in the last few years is the use of reputation information for network entities (servers or domain names). The idea is that if a client attempts to contact a domain or server with poor reputation it should be stopped, since that will stop also its exposure to potential malicious activity. Reputation data is often compiled into blacklists, i.e., list of domains and IPs that should be avoided, and distributed to devices that enforce the blocking of elements on the blacklist. For example, devices that may use reputation data include firewalls, proxies, and URL filters.
Malware authors have a crude but effective attack against such reputation blacklists: they can use a certain server or domain for malicious purposes only for a very limited amount of time. After its IP or domain name is "tainted", that is, has entered one or more blacklists, it is simply abandoned and no longer used. This strategy imposes additional effort and expenses on the attackers (they need to register new domain names or manage new servers with high frequency), but it is effective.
Recent data from researchers at Google shows that this strategy is in fact already well in use: they studied domains hosting exploit kits used in drive-by-downloads and found that their median lifetime is only 2.5 hours [41]. Clearly, an effective blacklist should be able to detect the malicious domain and distribute this knowledge to all the enforcement devices before the domain has been abandoned.

# The Command and Control Problem

## C.2 Command and Control Activity

We have seen that today's attacks are targeted, evasive, and aim at obtaining and exfiltrating sensitive data. How are these attacks carried out in practice? While the specific attack steps and their naming may vary across publications [19, 55, 73], the literature agrees on the general structure of targeted attacks, which is commonly represented as a sequence of steps similar to those of Figure 1.

## Reconnaissance

This is where the attacker learns more about its target and identifies the weaknesses that will be exploited during the actual attack. The reconnaissance activity encompasses both computer systems and individuals. Attackers examine their target's networks and systems by using traditional methodologies, such as port scanning and service enumeration, in search of vulnerabilities and misconfiguration that could provide an entry point in the organisation. Attackers also collect information about key people in the targeted organisation, for example by combing through data available on social media websites: this information will be used to facilitate later stages of the attack.

## FIGURE 1: TARGETED ATTACK LIFE CYCLE

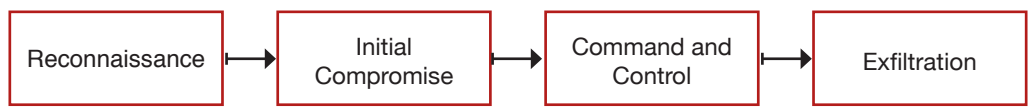

Reconnaissance → Initial Compromise → Command and Control → Exfiltration

## Initial compromise

This stage represents the actual intrusion, in which attackers manage to penetrate the target's network. Most frequently, the method of compromise is spear phishing. A spear phishing message may contain a malicious attachment or a link to a malicious web site [125]. Often times, the content of the spear phishing message are tailored based on the information acquired during the reconnaissance stage, so that they appear credible and legitimate.
A second common method of intrusion is the strategic compromise of websites of interest to the victim (or "watering hole" attack). In these attacks, attackers place mali- cious code on sites that are likely to be visited by the intended target: when the target visits the compromised website, she will be exposed to one or more exploits. Watering hole attacks represent an evolution of the traditional, opportunistic drive-by-download attacks [95, 97], in which victims are attracted, by different means, to a malicious web page. The web page contains code, typically written in the JavaScript language, that exploits vulnerabilities in the user's browser or in the browser's plugins. If successful, the exploit downloads malware on the victim's machine, which as a consequence, becomes fully under the control of the attacker [92, 96].

## Command & Control

The Command & Control phase of the attack is the stage where adversaries leverage the compromise of a system. More precisely, compromised systems are forced to establish a communication channel back to the adversary through which they can be directly controlled. The C2 channel enables an attacker to establish a "hands-on-keyboard" presence on the infected system (via so-called remote access tools), to install additional specialised malware modules, and to perform additional malicious actions (e.g., spread to other machines or start a denial of service attack).

## Exfiltration

In this stage, the attackers extract, collect, and encrypt information stolen from the victim's environment. The information is then sent to the attackers, commonly through the same C2 channel that was established earlier.
Of course, the exfiltration of data has been a key step in opportunistic attacks as well and it has been well documented in the literature. For example, studies of the data stolen (or "dropped") by the key loggers components employed in banking trojans have reported on the amount of data being transferred, its estimated value, and the modus operandi of their operators [50]. Furthermore, researchers have sinkholed or hijacked entire botnets with the goal of gaining an inside view of the data stolen from infected machines and the operations of botmasters [114]. With respect to the exfiltration techniques seen in these traditional attacks, we expect targeted malware to expand more effort into disguising its exfiltration activity and its infrastructure.

# The Command and Control Problem

## Differences with other models

In this exposition, we have simplified the attack models discussed in the literature, to avoid distracting the attention from the main purpose of this study: the Command & Control phase. More detailed descriptions of other phases may be useful for readers focusing on other steps of the attack chain, such as the initial compromise.

In particular, Hutchins et al. [55] emphasise the steps required to perform the initial intrusion by introducing specific phases named Weaponization, Delivery, Exploitation, and Installation. Since we focus on the C2 stage, we group all these phases under the generic Initial Intrusion label.

Mandiant's report [73] emphasises instead the steps performed by attackers after the initial compromise and leading to a persistent presence inside a target's network. After a stage named Establish Foothold, the authors present a cycle of steps (Escalate Privileges, Internal Recon, Move Laterally, and Maintain Presence) that enable attackers to establish an expanded foothold inside the target's network. The authors point out that these steps are optional and may not occur in all attacks. We consider these activities to be part of the compromise phase.

## C.3 Statistics

It is notoriously hard to obtain adequate statistics on information security in general, and to measure the volume and impact of cyber attacks in particular. As a high-profile example of the difficulties of this task, a recent review has found significant flaws in a report [27] commissioned by the UK Cabinet Office from Detica, which provided high estimates for cybercrime's annual cost to the UK [4].

The sources of data on cyber attacks have traditionally been surveys and telemetry collected by security vendors across their installation base. Both sources have their own issues [5]. For example, surveys often introduce bias by collecting most of their responses from large companies, which have the resources to collect the data requested by surveyors. In turn, statistics from security vendors have often been scrutinised for issues of over-reporting, which increases the perception of the risk involved with security threats and potentially favours the sales of a vendor's product. To further complicate the matter, conclusions from different sources have sometimes been found to be significantly different, if not contrasting.

Recent legal developments may help the collection of meaningful security metrics: in the last few years, disclosure laws have been introduced requiring businesses to report security incidents involving the theft of personal data [15]. Such regulation may increase the collection of attack metrics available, but at the moment they only cover specific incident types and geographic areas. Collecting sound statistics for targeted attacks seems especially challenging: this activity shares some of the same problems found with quantifying cyber attacks in general and it adds a few issues that are specific of this particular domain:

• The victims of targeted attacks are likely not willing to disclose the information that they have been attacked or, worse, breached. This knowledge may be embarrassing with customers and regulatory agencies, and the disclosure details may provide useful information to competitors (e.g., information about new product lines). Even more problematically, as we will see, targets may not know for a long time that they have been attacked.

• Targeted attacks span vast sectors of the economy, therefor, it may be difficult for any single entity (security vendor or governmental agency) to have a sufficiently broad visibility of the problem.

• Different reporters may have different definitions for what counts as a targeted attack and for the reporting methodology. It is not unusual to encounter descriptions of targeted attacks from one vendor that other vendors classify as traditional attacks.

Even with these caveats in mind, we present here a number of statistics from different sources. We consider reports that provide (aggregated) data about targeted attacks. We do not, instead, include in our review reports that only describe the attacks in general. For the reasons we have discussed, the statistics reported here should be approached with a healthy dose of caution, in particular with regard to their ability to support general inference about targeted attacks; but they still provide a snapshot, an initial quantitative look into targeted attacks, offering some light on questions such as their pervasiveness and their usual targets. We hope that in the future more and better data on targeted attacks will be available, enabling more robust quantitative analysis of this phenomenon.

## Mandiant Report

Mandiant is a security vendor providing incident management products and services to large institutions. Due to their business focus, Mandiant has built a reputation of dealing with targeted attacks. They publish a yearly report with findings from their engagements; the latest available report covers data from 2012 [72].

Mandiant's report suffers from some of the general problems we have discussed earlier. In particular, the sample size (the number of incidents) used as the basis of the report is not specified. Similarly, it is not clear if the incidents analysed in the report (those affecting Mandiant's customers) are a sample set representative of the general population. Nonetheless, the report contains a number of statistics that are worth discussing, as they offer an initial characterisation of targeted attacks. First, it discloses that only 37% of the intrusions were discovered by the victim itself: in the remaining cases, the victim was notified by some external party (e.g., law enforcement, customers, security vendor). The median time during which attackers are able to maintain a presence in the intruded network is reported to 243 days, well over 8 months. They also report that in 38% of the cases, attacks are repeated, supporting the notion that attackers are persistent once they have identified an intended target.

The list of targeted sectors include: aerospace and defence (17% of the cases), energy, oil, and gas (14%), finance (11%), computer software

# The Command and Control Problem

and hardware (8%), legal (7%), media (7%), telecommunications (6%), pharmaceutical (4%), other (25%). The report does not elaborate on how the business classification was drawn.

## Symantec

Symantec is a large computer security company, focusing on virus protection. They publish a yearly report on the status of Internet security; the latest available data covers the year 2012 [120]. The report investigates targeted attacks on the basis of the targeted malicious emails identified by Symantec's products. In total, the analysed dataset comprises about 55,000 attacks. The methodology used to discriminate whether a malicious email is targeted or opportunistic is intuitively presented, but there is no detailed description of the algorithm used to make this determination.

Out of this dataset, Symantec reports observing a number of targeted attacks per day ranging from 50 to about 225. The report warns that one large attack campaign in April against a single target would have significantly skewed results and, thus, has been removed from the presented results: while a reasonable course of action, this observation questions the sample size and generalizability of reported data.

Also in this case, the report lists the targeted sectors: manufacturing (24% of the cases), finance insurance and real estate (19%), other services (17%), government (12%), energy/utilities (10%), services professional (8%), aerospace (2%), retail (2%), wholesale (2%), and transportation (1%).

The report also comments on the size of the targets: 50% of the attacks targeted large organisations (those with 2,501 employees or more), 31% small and medium business up to 250 employees. The analysis of the malicious email dataset also provides some insight into the targets of the initial compromise: R&D personnel (27% of the attacks), sales (24%), C-level executives (17%), and shared inboxes (13%). Interestingly, the report points out a handful of significant changes from previous year data (for example, R&D personnel used to be targeted only 9% of the cases in

2011 compared with 27% of 2012): it is not clear if these changes are an artefact of the data collection and analysis process or correspond to actual changes in the tactics of attackers.

## Verizon

The telecommunication company Verizon publishes a yearly report on data breaches. The last available report at the times of writing covers 2012 and contains data compiled from 19 organisations for a total of of more than 47,000 security incidents [127].

The report has a more general scope than those discussed so far (it covers data breaches in general), but it does provide some useful insights on targeted attacks. The report is characterised by a careful methodology, which is explained in detail.

The main findings relevant to our analysis is that 25% of the breaches they report on are targeted. The report confirms the elusive nature of attacks (targeted or not): 69% of breaches were spotted by an external party (9% customers), and 66% of the breaches took months or even years to discover. Another interesting observation is that, in most cases, the initial compromise does not require sophisticated techniques; in 68% of the cases it is rated as "low difficulty" and less than 1% as "high". More worryingly, subsequent actions may be more sophisticated: 21% are high and 71% are low. Unfortunately, the report does not break down these statistics between targeted and opportunistic attacks.

## Discussion

As we have anticipated, the available data is unfortunately somewhat limited and the reporting methodology used to analyse it is not always sufficiently described. This limits our ability to make generalizations on the basis of the data sources that we have briefly listed here. However, there are several points that are worth discussing, keeping in mind our objective of designing and deploying better defences against targeted attacks.

### Ability to detect

There seems to be support to the notion that attacks in general, and especially targeted ones, remain unnoticed for a long time. This indicates that organisations do not have appropriate controls, tools, and processes to identify the presence of intruders in their network. Unfortunately, we do not possess enough data to conclusively point to the precise reasons for why this occurs: in particular, it may be result of factors ranging from cultural ones, such as the lack of appropriate awareness to this specific risk (the "I'm not a target" mentality), to technological reasons, such as the unavailability (real or perceived) of effective defensive tools.

The failure to detect intrusions for months, if not years, also implies that attackers have a long time to carry out their attacks, compounding the damage inflicted on the target organisation. At the same time, from a defensive point of view, it shows an imbalance between two common defensive strategies targeting different steps in the attack kill chain: defending by preventing the intrusion and defending by detecting an intruder. More precisely, detecting the initial compromise requires to catch and identify the individual event that leads to the intrusion (e.g., the receipt of a specific spear phishing email or the visit of a specific malicious web site). This is an event that occurs in a specific point of time, and, lacking forensics capabilities, its detection requires that at that specific time some defensive tool (e.g., intrusion detection system or anti-virus tool) is capable of performing the detection. Intuitively, detecting the presence of an intruder may instead happen at any time after the intruder has established a presence in the target's network, during which time defensive tools may be updated or improved. This situation flips the imbalance between attack and defence: while detecting the initial compromise requires that defence tools work effectively at all times (a single missed detection may lead to the compromise), to detect the intruder's presence it is sufficient that the attacker makes a single mistake, revealing its presence.

# The Command and Control Problem

## Targets

Different sources provide different lists of targeted sectors. This could certainly be a result of the selection biases inherent in each source. However, what we can conclude by looking at the target lists provided, for example, by Mandiant and Symantec, is that any significant sector of the economy may be the target of attacks. Also relevant is the observation that an organisation size is not a predictor for being attacked or not: organisations of any size, from small businesses to large corporations, have been attacked in the past.
In our context, this observation has important consequences. From an organisation point a view, the results and recommendations that we provide in this report should be generally applicable. From a vendor perspective, this points to the need of providing tools and mechanisms that are amenable to widely different organisations, with large ranges of technical skills, human resources, and budget numbers.

## Sophistication

Third, the Verizon report contains some preliminary data about the technical sophistication of attacks and, more precisely, it finds that a majority of the initial compromises are carried out with low difficulty techniques. While it is not clear if the same results hold when considering only targeted attacks (rather than looking at all kinds of attacks including opportunistic ones), this observation does match anecdotal experience from individual attacks, which often do not show particular sophistication, such as 0-day exploits or stolen digital certificates.
The lesson learned from this data point could be that, while defending against sophisticated attacks is becoming increasingly necessary, one cannot discard traditional attack techniques.

## C.4 Case Studies

In this section, we review a number of cases of targeted attacks. There exist many more accounts of attacks than we can report here; we decided to focus on those cases that show particular attack techniques and objectives, or that highlight the importance of C2 detection. More precisely, we include a set of cases that have been publicly disclosed: these attacks affected high-profile organisations and had egregious impact. We also include a set of cases that were provided by Lastline, a security company providing solutions to defend against advanced attacks. These cases, opportunely anonymised to protect the identity of the targeted organisations, are based on Lastline experience "in the trenches", working with its customers, and draw attention to specific aspects of attacks or to problems with existing security approaches.

## Political espionage

In January 2013, the New York Times publicly denounced that it had been subjected to targeted attacks for a period of four months. The attacks had been traced to Chinese hackers and were linked to an ongoing investigation at the journal that was highly critical of the Chinese political elite [91]. Further investigation of the attack methods and objectives linked the attack to a larger attack campaign targeting news and media companies, including Bloomberg News, which was compromised the previous year.
An investigation of the incident found that the attack activity showed some of the traits typical of targeted attacks originating from China: attackers hopped through com- promised accounts at US Universities, as a way to hide their identity and make investigation more complex, and they were suspected of using spear phishing to gain the initial access to the Times' network.
The following steps of the attack fully reveal the targeted nature of the incident: the attackers obtained the passwords for every Times employees and used them to gain access to the personal computers of 53 of them. Then, they deployed code to search for and steal documents kept by reporters on the current investigation on Chinese politicians.
The Times article contains two pieces of information that are useful to illustrate the limitations of traditional security approaches that are based on the detection of the initial intrusion activity. The article comments that the spear phishing attack completely bypassed existing defences at the perimeter: "Attackers no longer go after our firewall. They go after individuals.".
The Times also reported that of the 45 malware samples used in the course of the intrusion, only one was identified by the journal's anti-virus tool, whose vendor later issued a statement reading "We encourage customers to be very aggressive in deploying solutions that offer a combined approach to security. Anti-virus software alone is not enough." [122].

## Military espionage

In May 2013, the confidential version of a report prepared by the Defence Science Board for the Pentagon was leaked to the Washington Post [83]. The report claimed that the designs for many of the US advanced weapons systems had been compromised by Chinese hackers. The report claimed that the extensive theft had targeted the documentation for several missile systems, combat aircraft, and ships.
While there are at the moment few details regarding how the intrusion actually occurred, it appears likely that the attacks targeted in particular large military contractors, which are involved in the design and production of the military systems.
This case study is a cogent example of attacks aiming at obtaining valuable intellectual property: sources from the Washington Post claimed the stolen designs were the result of 15 years worth of research and development.

### Supply chain attacks

In February 2013, the security vendor Bit9 reported that it had been compromised [67]. Bit9 produces a whitelisting product, which specifies the list of software that should be allowed to run in a network; anything else is considered to be dangerous. As a consequence of the intrusion, the attackers managed to steal the secret certificate that Bit9 uses to sign its software releases. The company also revealed that some if its customers had received malware that was digitally signed with the stolen certificate. Also in this case, the specific details

# The Command and Control Problem

of the intrusion are not entirely clear. There are, however, several interesting aspects in this attack. First, attackers compromised Bit9 with the primary intent of acquiring the capability required to successfully attack upstream targets protected by the company's products. These are often called "supply chain attacks" because they target one link of a chain that eventually leads to the organisation that is actually being targeted.

## Manufacturing espionage

Military secrets are hardly the only ones to be sought after by attackers. In early 2013, Lastline started monitoring the network of a manufacturer active in the field of fashion. During this monitoring, it determined that an internal server was infected: further investigation revealed an unexpected remote connection to that server originating from China. Among other data, the server contained all the designs of the manufacturer's new collection, which had not been officially presented.
This episode shows that the data targeted by sophisticated attackers is not limited to highly confidential documents. Industrial espionage, in certain cases conducted with semi-official governmental blessing, has targeted a large spectrum of economic sectors [73].

## Malicious infrastructure agility

In mid 2013, a Lastline product was installed at a professional services firm. Lastline detected a successful drive-by-download attack against one of the firm's employees. The drive-by had started when the employee visited a legitimate web site that had been compromised; the web site collects information relevant to the firm's business. Additional drive-by-download attacks were detected short thereafter, originating again from web sites belonging to companies and organisations in the same business field as the firm.
After one of the successful drive-by-download attacks, connection attempts to a malicious domain were observed: the connections did not succeed because the destination server failed to respond. However, a day later, connections to the same domain were observed: this time, the server was responding and was actually distributing the configuration file of a widespread financial malware.
These events suggest that attackers run campaigns targeting specific business sectors (these could be considered less targeted versions of the watering hole attacks). Furthermore, the sudden activation of malicious domains shows that the malicious infrastructure used by attackers (exploit sites and C2 domains) can vary quite rapidly, thus requiring its constant and up-to-date monitoring.

## Built-in polymorphism

This case study was collected after the installation of Lastline product in a University environment. Here, an administrative user received a malicious email and clicked on link contained therein twice in a short span of time. The link caused the download of a malware program. Interestingly, the binaries downloaded as a consequence of the user's actions were different: they not only had different hashes, but they also received different scores on VirusTotal, an online service that scans submitted binaries with over 40 anti-virus tools.
This episode shows that the use of evasion techniques, polymorphism in this case, is a built-in component in many attacks: all the binaries downloaded in the course of the attack are (superficially) different. In addition, this case illustrates the importance of user security education: users are all too often a weak link in the security of an organisation.

## C.5 A New Focus

There are several lessons that we can learn from the statistics on today's attacks and the case studies that we have presented.
One is that *preventing compromises may be difficult*. We have seen that intrusions happen, even at security conscious organisations which possess considerable domain knowledge, expertise, and budget for security. We have also seen that there does not really appear to be a sector that is immune from attacks: modern businesses and organisations handle on a regular basis confidential data of various nature (personal, financial, R&D) that is valuable to attackers. It should also be noted that the compromise may initiate outside of an organisation's perimeter (and away from the defences that the organisation has put in place), and then spread inside it as the infected device re-enters the perimeter. For example, with bring-your-own-device (BYOD) policies, organisations explicitly allow employees to bring on the workplace personally-owned and managed mobile devices, such as smartphones, and to use these devices to store privileged data and to interact with internal systems.
As a consequence, it is clear that *detecting that a compromise has occurred is critical*. Ideally, the detection is performed as early as possible in the life cycle of the attack, to limit the damage that is suffered (for example, before confidential data is actually stolen). Unfortunately, we have seen, in particular with the Verizon data on breaches, that a significant number of infections go completely unnoticed for a long amount of time.
C2 detection and disruption seems to offer a solution to this problem: by focusing on the C2 phase of an attack, one accepts that a device may become under control of attackers, may enter organisation's network, and may even acquire confidential data. However, the successful detection of C2 activity will preclude the attackers from performing the actual malicious, damaging activity of their actions, such as stealing confidential data. Of course, C2 detection should be seen as a complementary approach to the prevention of compromise, rather than a substitution for it: completely blocking an attacker, whenever possible, is preferable than having to deal with it after the fact. With this approach in mind, we review in the next sections the techniques that attackers use to set up C2 channels, and then the approaches that have been proposed to detect and disrupt such channels.

# C&C Techniques

As we have already discussed there is a constant battle between the attackers (malware writers) and the defenders (security professionals), wherein the defenders find a new way to detect and block attackers, and in response the attackers come up with new, often novel ways of performing their C&C communication to evade the defenders. In this section we will discuss various techniques used by the attackers, including some in-depth case studies of actual malware that use them, and describe the general trends that the malware is exhibiting.

The command and control system for most modern malware has three components. These are **controller discovery, bot-controller communication** protocol and the **C&C topology**. In the controller discovery phase, the malware attempts to identify the location of the control system. The topology of the system may take many forms, falling into the broad categories of centralised and de-centralised. Finally, there is the actual method of communication from the malware to the controller. These three steps are often completely separated, and it is a common occurrence for malware to update one of these components while keeping the other components constant. This section is structured as follows: first, we will give a brief insight into the trends in malware command and control over the years. We will then describe the various techniques used by malware to perform the three actions as described above.

## D.1 Overview

Over the years, the architecture of the C2 channel has evolved substantially, driven by an arms race with detection-response mechanisms. The network structure (or topology) of the C2 channel has an intimate relationship with its resilience to attack and error, as well as scalability to larger numbers. C2 designers desire scalability, robustness to take-down efforts, and stealth – anonymity against detection.

## C2 communication structure

Early C2 designs followed a centralised architecture such as using an IRC channel. In this design, administration and management tasks are simple and the architecture tolerates random losses of C2 nodes with little impact on efficiency. However, such a topology is fragile against targeted attacks — if the defenders can identify the channel and attack or take down the server, they effectively disable the C2 channel. Such fragile architectures were accompanied by poor software engineering practices. For instance, the address of this server was often hard-coded in to the malware and static in nature.

However, the growing size of botnets, as well as the development of mechanisms that detect centralised command-and-control servers [10,12,40,43–45,63,71,117,135], has motivated the design of decentralised peer-to-peer botnets. Several recently discovered botnets, such as Storm, Peacomm, and Conficker, have adopted the use of structured overlay networks [93,94,116]. These networks are a product of research into efficient communication structures and offer a number of benefits. Their lack of centralization means a botnet herder can join and control at any place, simplifying ability to evade discovery. The topologies themselves provide low delay any-to-any communication and low control overhead to maintain the structure. Further, structured overlay mechanisms are designed to remain robust in the face of churn [47, 70], an important concern for botnets, where individual machines may be frequently disinfected or simply turned off for the night. Finally, structured overlay networks also have protection mechanisms against active attacks [16]. Fully decentralised topologies offer systematic resilience guarantees against targeted attacks on the C2 channel, yielding new forms of robustness. The vast power of peer-to-peer botnets from the use of resilient topologies comes at the cost of stealth; the unique structure can be also used as a point of detection [82].

## C2 communication traffic

*Relationship between traffic and structure.* C2 evolution has also been driven by largescale defence efforts to isolate C2 traffic based on its unique traffic characteristics. Defence efforts to block unused ports and application protocols simply encouraged C2 designers to tunnel their traffic through legitimate services, paving the way for the return of centralised architectures – C2 channels has been observed in the wild using comments in HTML pages or even actual blog posts on public forums to communicate. Since traffic is routed through legitimate services, defenders are effectively denied the option of disabling the service, as doing so would hurt legitimate interests.

*Communication traffic-pattern anonymity.* Other drivers of C2 techniques have been defence efforts arising from the application of statistical traffic analysis techniques. These range from simple anomaly detection techniques to sophisticated machine learning based detection. In response, C2 designers have adopted evasion techniques [87, 107] to hide traffic patterns from detectors. Such techniques mask the statistical characteristics of C2 traffic by embedding it within synthetic, encrypted, cover traffic. The adoption of such schemes only require minimal alterations to pre-existing architectures and can be adopted on a strap-on basis by other C2 operators. We can expect these techniques to mature quite well in the near future.

*Communication end-point anonymity.* An interesting development is that C2 designers have adopted anonymous communication techniques within their architectures. Initial designs used simple anonymous proxies or stepping stones to route the traffic through a number of proxies to anonymise C2 traffic endpoints. More recently, C2 designers have started abusing systems designed for Internet privacy such as Tor, JAP, and anonymiser [30, 59]. End-point anonymity prevents defenders from isolating (and filtering) C2 hosts even if they can successfully detect traffic patterns.

*Communication unobservability.* C2 techniques in the form of practical unobservable communications have also been developed. These are especially powerful as they offer full unobservability; the strongest possible anonymity guarantee, subsuming both

# C&C Techniques

end-point anonymity as well as traffic-flow anonymity. Currently adopted techniques are based on the use of covert communication techniques. For instance, the application of probablistic information-hiding techniques such as image sharing behaviour on social networks [81]. Emerging trends include, the use of uncompromised DNS servers as transient stores of C2 payloads. Techniques to evade responses. The primary response mechanism against C2 channel is to isolate domain names and IP addresses related to C2 activity. In response, C2 architects have developed techniques inspired by fault-tolerance literature. This is characterised by the evolution of domain generation algorithms (DGAs), and fast-flux networks which can allow large numbers of IP addresses to be linked to a single domain.

## D.2 Communication structure

### Centralised architectures

Early C2 designs were based on a centralised architecture where one or more servers are exclusively used to coordinate C2 communication. The classic design for command and control in malware is to make use of an Internet Relay Chat (IRC) server. IRC was developed in 1988, and is a protocol used for text chat over the internet. Its primary function is to provide "channels", which are chat rooms allowing for group conversations (private user-to-user chat is also available but less common). Channels are hosted on servers, which in turn are part of IRC networks. While most channels are publicly accessible, it is possible to require authorisation to join a channel. User within a channel have varying levels of access (modes), which can define what that user can do within a channel. The channel itself has modes which define what each user mode can do, such as changing the channel topic, and other options regarding the channel (such as access authentication). All of this makes it a simple platform for malware communication. It provides a simple function for the attacker to deliver commands to the malware, and it's equally simple for the malware to transmit information, such as collected data, back to the human controller. Centralised architectures are simple and easy to manage, and robust to the failure of large numbers of malware-infected computers. In 2000, researchers [3] famously showed that while centralised architectures are robust to random failure, they are fragile against strategic attacks; removing the high-centrality components of the communication structure disables the C2.
Further, centralised C2 networks are not scalable. Supporting a large C2 network consisting of hundreds of thousands to millions of malware instances requires careful coordination amongst a large number of control servers, each servicing a few thousand or so malware instances.

### Decentralised architectures

To counter the structural weaknesses and scalability limits of centralised architectures, many C2 designers are moving to decentralised or peer-to-peer (P2P) architectures for command and control. The main design goals of these architectures are: scalability (nodes maintain a limited state and communication costs grow slower with the number of nodes), fault tolerance (requests can be routed around failed/takedown nodes) and P2P nature (distributed architecture with no single point of failure and strong availability guarantees).
In a P2P network, there is no central control server; instead every member of the network acts (or can act) as a server, thus providing a load balancing property. Further, decentralisation ensures large amounts of redundancy against targeted attacks, consequently in comparison to centralised C2, any takedown effort will need to attack a significantly larger percentage of the nodes to completely disable the C2 network.
The use of decentralised C2 networks has heavily borrowed from P2P file sharing networks (used for both legal and illegal means). In a P2P network, each member communicates to a non-uniform number of other members or 'neighbours'. Nodes only communicate with their neighbours in the network, with different variants of P2P networks providing different methods for the routing of data around the network. P2P networks can either be unstructured overlay networks ( Bittorrent, Gnutella, or Kazaa). Or, structured overlay networks such as CAN, Chord, Pastry, deBruijn-based options (Koorde, ODRI, Broose, D2B), Kautz, Accordion, Tapestry, Bamboo, and Kademilia. We have named a few but there are many other options, which indicates the substantive depth of the design possibilities.
We will now take look at the typical operation of the Bittorrent network. To access a file on the network, the user downloads a tracker file, which contains a list of peers that hold some or all 'pieces' of the file. The user then directly connects to these peers, and downloads the pieces they have. Eventually, you will have the entire file. The more users who are involved in the "seeding" (hosting) of file pieces, the faster the download speed. Obviously, currently the most common use for this technology is in the sharing of (often copyright) media files. The network can, though, provide an easy method for propagating information among a large number of users without the use of a central server. The typical situation for malware is that the malware will have a list of peers to which they are connected, and they will repeatedly check with these peers for new commands. The bot controller simply has to "upload" the commands to a single (or group of) nodes (which can be anywhere in the network), and the command will eventually reach all nodes through a flooding mechanism. This has the additional advantage that there is no need for a link between the data and the uploader, as is the case with a centralised system.

### Case Study: Storm

Storm is an good example of a botnet that uses a p2p network for its command and control. The Storm botnet, at its peak in 2007, comprised of anywhere between 1 and 50 **million** infected hosts. Storm propagates solely through the use of spam emails, which contain links to either websites that take advantage of browser exploits, or prompt the download of malicious software. One of the first actions the malware performs is to make sure that the system clock is correct. This is vital for communication. Storm makes use of OVER- NET, which is a Kademlia based distributed hast

# C&C Techniques

table (DHT) based p2p network. Each bot has an 128 bit DHT id, which is randomly generated. Routing is performed by computing the XOR distances of the IDs to the destination, node a which has a message for node d will forward to the peer (neighbour) with the closes id to d. Storm, like many p2p networks, uses a publish/subscribe style of communication. A node publishes information using an identifier generated from the contents of the information. Consumers can then subscribe to the information using the identifier. The bots compute identifiers to subscribe to using the day and a random number between 0 and 31. The controller can precompute these identifiers and publish information using them. The information published consists of a filename of the form "*.mpg;size=*", where * represents a 16 bit number. The malware converts this to an IP address and port number, at which point the malware performs a direct connection to talk to the controller directly.

## Social Networks

Social networks now play a huge part in many peoples lives. The benefits that they bring to both businesses and end users are hard to ignore. In fact, Facebook, the largest social network, now has over **1.1 billion** users, and is currently the second most visited website (www.alexa.com) in the world. The sheer volume of social network traffic, plus the ability to easily host information within a social network page for little to no cost, has made them a very attractive tool to malware creators. In this case, C2 channels are built as an overlay network over a social network, again both centralised and decentralised configurations are possible. Although social networks are largely based around a small number of highly well-connected central servers, it's not possible to simply block the OSN due to their immense popularity with legitimate users. Further, online social network (OSN) providers have invested significantly in computing infrastructure. OSNs feature worldwide availability and load balancing, thus mitigating the traditional scalability limits of centralised C2 channels. They host a rich variety of content enabling the use of steganographic communication techniques, which we will discuss this in future sections of the report.

There have already been numerous examples in the wild of malware that uses social networks (or similar sites) as part or all of the command and control system. One (possible) botnet that has been found is an unnamed piece of malware that receives its commands through tweets posted to a particular Twitter account [132]. It is unclear however if in this case this was a researcher testing a new toolkit for Twitter command and control rather than an actual botnet. An example found by Arbor Networks [85] also demonstrates a botnet using Twitter as part of it's C2 channel; in this case the twitter account posts base64 encoded URLs, which represent secondary C2 servers. The same behaviour is also found on identically named Jaiku and Tumblr profiles. They also found a botnet that uses a malicious application on the Google App Engine cloud hosting platform which also returns URLs to which the malware will then proceed to connect to [84].

A confirmed piece of highly targeted malware that is using a social network as part of its command and control is Taidoor. Taidoor attacks organisations that have links to Taiwan (hence the name), and security firm FireEye have found that the malware has been modified to host the actual malware binaries in a Yahoo blog post [129]. The malware is initially delivered by email and performs an exploit against Microsoft Office, and a downloader is installed. This downloader connects to a Yahoo blog post, which contains seemingly random data. The data is in fact the actual malware binary, contained between two markers and encrypted using the RC4 stream cipher, with the resulting cipher-text being base64 encoded. When decrypted, the data is a dll file containing the malware. The malware then connects directly to two C&C servers.

## D.3 C2 communication traffic

Communication pattern analysis. A key

feature of C2 channels that distinguishes them from other malicious activity is the fact that the individual malware-infested hosts communicate with each other. This lets them carry out sophisticated coordinated activities, but it can also be used as a point of detection. Traffic analysis techniques can be used to detect communication patterns among bots, and how such patterns can be used for more effective botnet countermeasures, and tradeoffs between botnet performance, resilience, and stealth. *Traffic analysis* is an old field hence many of the techniques are applicable. We will review these in Section E. Consequently, C2 designers have adopted techniques to hide communication patterns. *Traffic analysis resistance,* is a much desired property by C2 designers. Anonymous communications technologies study the design of communication channels that are resistant to traffic analysis. For C2 designers, the ultimate goal is unobservable communications. The property of *unobservability* – the strongest form of communication anonymity – refers to the communication capability that a third party cannot distinguish between a communicating and non-communicating entity. For instance, by appearing indistinguishable from legitimate traffic, C2 activity will be effectively undetectable. There's a substantial amount of knowledge in the public domain on the topic on which C2 designers can build upon.

We now briefly review the current state of adoption of anonymous communications technology by C2 designers.

## Tor

Tor (originally TOR:The Onion Router) is a service used to provide anonymity over the internet. It is used by governments and the public alike (for example it is extremely popular with whistle-blowers), and even receives a large proportion of its funding from the US Department of Defence. The basic system works by relaying internet traffic through a number of nodes, and applies multiple levels of encryption/decryption to mutate the traffic after each hop. It is extremely difficult to identify the original sender and receiver of packets sent over the network. This security has also made it

# C&C Techniques

a target for malware coders, and there have been cases of malware that use the Tor network (and some of its extra features) to aid in command and control. To become a part of the Tor network, one simply has to install a simple piece of software. The machine can then act as a replay node for others, and make use of the Tor network.

One of the more advanced features of Tor is the ability to set up Hidden Services. These allow a server to hide behind a proxy, keeping the actual identity of the server hidden from those who access it. Hidden services work by setting up "Rendezvous" points. A rendezvous point is a node on the Tor network, whicis used as the entry point for the server. Traffic between the rendezvous point and the server is routed in the normal Tor fashion, providing anonymity. A rendezvous point is access using an ".onion" link.

While few examples of actual bots have been identified that use Tor as part of their C&C channel, there is growing evidence that this is occurring on a large scale. In late August/early September 2013 the Tor network experienced a large increase in the number of users [29]. The actual amount of traffic on exit nodes, however, only showed a minimal increase. This was eventually identified to be down to the SDC botnet [134]. The SDC botnet hosts its command and control server behind a Tor hidden service. The botnet, however, shows little activity, and is believed to simply be used for installing other malware.

## Case Study: Skynet

Skynet is a moderately-sized (~12000 machines) botnet based upon the Zeus family of malware. The interesting thing (apart from the usage of Tor) about Skynet is that its operator hosted an IAmA (Q&A) session on Reddit 1. When a team of researchers [46] discovered an instance of the malware, they were able to use the information provided by the Reddit post, plus a small amount of reverse engineering, to provide an almost complete profile on the operation of the botnet.

The malware is spread primarily through the Usenet file sharing network, and is primarily used for DDoS attacks, data mining and Bitcoin mining. When the malware is installed onto a machine, the Tor client for Windows is also installed, and a Tor hidden service is set up for the machine itself. All C&C communication is performed over a Tor SOCKS proxy running locally on the machine. The hidden service is opened on port 55080. The primary method of C&C is an IRC server hosted behind a Tor hidden service. The server runs at "uy5t7cus7dptkchs.onion" on port 16667. The controller issues com- mands to the malware through the IRC channel. These actions can include performing attacks and returning info on the host machines.

The malware also includes a version of the Zeus malware family. Zeus is a very common banking trojan, with a primary goal of stealing personal financial details (for example credit card numbers and online banking passwords). Zeus provides a web- based C&C server, which the controller has hidden behind a second Tor hidden service. By accessing the control server, the researchers were able to recover a XML file con- taining the current target websites.

The final component of the malware performs Bitcoin mining. The malware includes the open source "CGMiner" software used for Bitcoin mining, which connects to a number of Bitcoin mining proxy servers. Interestingly, seven IP addresses for proxy servers were found, of which two were active, but none were hidden by Tor.

Due to the use of Tor, it is almost impossible to identify the actual location (and owner) of the command and control servers. Through responses on the Reddit post, plus the botnets concentration in central Europe (in particular the Netherlands and Germany) there is a strong chance that the operator is based in Germany.

## Unobservable C2 Communications

Whereas systems such as Tor aim to provide anonymity through unlinkability (i.e. disguising who is talking to whom), unobservable communication methods aim to hide the fact that anyone is communicating altogether. Tor, and similar systems, are designed to provide low-latency communications but this is often difficult to achieve when using unobservable communication methods, which often provide a higher-latency for of communication. While this is often deemed unacceptable for the user-base of Tor like systems where usability is a factor, it is not an issue for malware coders. The most common form of providing unobservable communications is through the use of steganography.

*1 http://www.reddit.com/r/IAmA/ comments/sq7cy/iama_a_malware_ coder_ and_botnet_operator_ama/*

Steganography (Greek: "concealed writing") is the art of writing messages in such a way that nobody, apart from the sender and receiver, suspects the existence of the message. Steganography is an art that has been used for thousands of years, and has been reinvigorated in the digital age. The main purpose of using steganography is that it can make the communication **unobservable**. There are two ways in which steganography can be used by malware to hide the command and control communications. The first is that the malware can make its communication protocol appear as another, and secondly it can embed itself within otherwise legitimate content online, such as images.

Today most media types, including text, images and video, are capable of contain- ing hidden data in a number of ways. In the simplest cases, this can be achieved by adding extra metadata to files to store the required information, although this is easily discovered. The alternative, and more advanced, method is to change the actual file contents itself. For example, in a image file the least significant bit of each pixel in the image can store the data. This will allow a relatively large amount of data to be stored (directly corresponding to the size of the image), and to the untrained eye the image will appear to be unchanged.

## Unobservable C2 Communications

In a audio file, the data can be hidden

# C&C Techniques

by introducing an echo, with the amount of delay indicating the data (the delay will be in the 10s of milliseconds so unobservable to the untrained ear). There are numerous algorithms for steganography, providing differing levels of unobservability and modification resistance (the classic way to remove image steganography, for example, is to resize or slightly distort the image). Currently, there are very few examples of steganography being used by malware in the wild. It is expected, however, that the amount of malware making use of steganography will increase as command and control detection methods become more advanced and difficult to circumvent using traditional means. Therefore, this section will mainly deal with proposed designs for steganography-based malware command and control.

One example of real world malware using a form of steganography is the Trojan.Downbot Trojan [137]. Trojan. Downbotspreads through targeted emails, and the first thing it does is to access an attacker-controlled website, for which the address is hardcoded into the malware itself. The website is made to look like a code tutorial site, and to anyone who happens to access them legitimately the website will appear completely harmless. If the page source is analysed, however, it can be found that the source contains specially formatted and encoded comments in HTML files, or extra bytes in image files. These comments and images contain the command and control commands for the malware, including a command for the malware to contact a specific IP/port combination (to upload collected data). This technique is effective as all the command and control communication is performed over HTTP and would appear in logs to be normal web browsing behaviour, and to block all HTTP traffic would cause usability issues for legitimate users.

## Case Study: Stegobot

Stegobot is a proposed design for a de-centralised botnet with an unobservable C2 protocol based upon steganography. The system utilises the existing user behaviour of the uploading of digital images to a social network, and the subsequent broadcast of these images to all of the users connections. The social network that is the focus of this paper is Facebook. The typical activity of Facebook is that a user uploads an image through the web interface, and while they are browsing the "news feed", the recently uploaded images of their connections are downloaded to a temporary folder on the local machine. The malware operates as a proxy - when an image is uploaded data is inserted into it before it is uploaded. The malware also attempts to extract data from the temporarily downloaded images, storing any recovered information. The malware is designed for information collection, in particular bank details and passwords. Commands are issued by the controller at some point in the network (note: this can be from any of the bots with with equal probability), and the command is then propagated through the use of flooding. Collected data is returned to the controller in the same way. The system uses a steganography algorithm that is hard to detect automatically (YASS), so provides reliable unobservability. The C2 system creates no extra web traffic as it uses exclusively the normal browsing habits of the affected users. When simulated on a social network of 7200 nodes, the bot controller can receive up to 86.13MB of data per month, which may seem like a small amount, but can represent many thousands of bank details and passwords.

## D.4 Evasion

## DNS

The Domain Name Service (DNS) is a naming system for computers on the Internet. It's a basic piece of infrastructure that translates human-relatable computer resource names to IP addresses which can be used to route information. Attackers extensively use it to build and operate the C2 channel.

### DNS Fast Flux

One of the major positive points for malware coders is that the IP addresses returned by an DNS request do not need to be static. This property is used by legitimate services through the use of Content Delivery Networks (CDNs). CDNs are used by large-scale web services (such as Amazon and Facebook) for a number of purposes, the primary purpose being to enable the use of multiple servers around the world, so a user can access content closest to them. It is also useful to aid in load balancing and provides extra redundancy in case of failure. In a CDN, a typical DNS response will contain multiple IP addresses, all of which are valid. The user will then connect to one of these. The returned IP addresses will typically have a TTL measured in hours or days. This is so that the effects of DNS caching can be felt; if the TTL is too short effective caching of results cannot be performed. Repeat request for the same domain from the same location will in general return the same set of IP addresses, possibly with some differences if the CDN needs to load balance.

An variation on a CDN is a Fast-Flux Service Network (FFSN) [51], in which the command and control server is hidden behind a wall of compromised machines, which are often part of a botnet. Each of these compromised hosts acts as a sort of "proxy" to the C&C server; each time they receive a request for the server they will forward it, and will return replies to the original requester. Each of these hosts will have a unique IP address, which can be used to access the server. For example, an FFSN comprising of 10,000 compromised machines provides up to 10,000 IP addresses that can all point to a single server.

The FFSN operates as follows. The domain of the server is public. A host wishing to contact the server makes a DNS request for the domain, and is returned a set of IP addresses, and then connects to one of them. This sounds familiar doesn't it? This part of a FFSN is almost identical to a CDN except for a few small differences. First, the returned IP addresses returned will not be for the actual controller(s), rather they will point to compromised machines within the flux network. Second, the returned IP addresses will have a very short TTL, measured in minutes (rather than the days as is the case in a CDN). A second request will in most cases return a

# C&C Techniques

completely different set of IP addresses. This is the fast-flux behaviour; the malware controller has no control over which of his compromised hosts are online so the returned IP addresses needs to change frequently to increase the risk that a hot is available.

### DNS as a Medium

It is also possible to use the DNS system as a communication channel rather than just as a way to set up the channel. One example of this being used in the wild is Feederbot [28]. Feederbot makes use of the fact that the RDATA field in a DNS response can be of multiple types, not just an IP address. One of these is TXT, which as the name suggests means actual text can be transmitted. Feederbot uses TXT replies to transmit data. The commands are encrypted and the encoded into base64 (which resembles random text). The remainder of the DNS response packet uses valid syntax, making detection difficult.

While Feederbot is optimised for one-way command and control, in most cases the malware will need to transmit information back to its controller. Seth Bromberger, working for the US Department of Energy, proposed a system for exfiltrating data from organisations by making use of DNS requests. In this case, the domain name that is being queried contains the data to be transmitted. The attack works as follows. The attacker sets up a domain name *(evil.com)* and makes sure that he has control of its authoritative nameserver *(nameserver.evil.com)*. Say, for example, an infected hosts wishes to transmit the data "Super Secret Stuff" back to its controller. It will simply make a DNS request for evil.com, pre-pending the data to the domain (so the request will be for *super.secret.stuff.evil.com*. The data can be encrypted before pre-pending to prevent the contents of the data being identified. When the request reaches the attacker controlled nameserver (nameserver.evil.com), the attacker can simply read off the data. The attacker can also send commands back to the malware in the response, either by using the method of Feederbot, or by, for example, using specific IP responses to indicate a particular task to be performed. A similar approach is proposed by Xu et al [133], who extend this idea to also include traffic patterns for the communication to avoid detection, for example by only creating DNS queries when the host machine is making them.

A second example of malware that makes use of the TXT record is the W32.Morto worm [79]. In this case, DNS requests to harcoded domains return encrypted binary signatures and IP addresses in TXT responses. The malware then downloads a binary from the included IP address. It is also rather peculiar in that there are no A records for the domains, only TXT records, indicating that the domains are used for the sole purpose of controlling the worm.

### Domain Generation Algorithms (DGA)

One method of providing resilience to both detection and reverse engineering is the use of Domain Generation Algorithms (DGAs). The function of a DGA is to allow the malware to programmatically generate domains for which it attempts to access a command and control server. It is then up to the attacker to ensure he controls the domains that will be generated.

A DGA will often be reliant on factors such as the current time or date, and the result should be consistent across multiple hosts. The malware will repeatedly run the algorithm to generate a domain and attempt to connect. The attacker can also run the algorithm, in advance, and register the domains when they are required to use as a temporary command and control server.

The main benefit to an attacker of a DGA is that they allow for a large amount of redundancy in the command and control server. The controller, at any one time, is short lived and so if one is taken down, a new one will be available in little time. For example, the Conficker malware will generate 250 domain names every three hours, based upon the current UTC date [94], The same domains are generated every three hours (8 times per day). The malware will do an lookup on every generated domain, and will attempt to contact every domain that has an assigned IP address to download binaries.

### Case Study: Torpig

Torpig is a botnet that is designed to steal personal information. In 2009, a team of researchers were able to take control of the botnet for a period of ten days, in which time they were able to document the operations of the botnet in its entirety [114, 115]. One of the key points of the Torpig botnet is that it makes use of a domain generation algorithm. Each bot independently uses a DGA to generate a set of domains based upon the current time. They then attempt to connect to each of these in turn, until one succeeds (i.e., the domain resolves to an IP address and the server replies with a valid response). The botmaster also computes the domains and registers them, usually with less than honest domain registrars, before they are generated, with the goal of getting at least one online. (The researchers were able to take control by beating the botmaster to it and registering the domains, effectively sinkholing the botnet).

### Future: Protocol Mimicking

One area of research that has grown recently is the area of protocol mimicking. The idea is to hide certain, noticeable communications by making them seem like they belong to a different protocol. The main area of focus on this so far is in obfuscating Tor traffic. In many cases it can be dangerous to use Tor, and it exhibits very noticeable communication patterns.

There are a number of systems that attempt to make Tor traffic appear as Skype traffic. As Skype is a widely used, low-latency and high bandwidth system, it is ideal to emulate. SkypeMorph [77], for example, attempts to make Tor traffic appear as a Skype video call. Both the client and the bridge node run the Skype client on a high numbered UDP port, and the client sends a Skype text message to the bridge containing its IP, UDP port and public key. The bridge replies with the same information. The client then starts a video call to the bridge, which it does not answer. Instead, the call is dropped and instead the encrypted data is sent

# C&C Techniques

over UDP between the ports opened for Skype. Once data communication starts, Skype is exited on both the client and bridge.

A slightly different approach is taken by StegoTorus [131]. In this system, Skype is not actually used, instead entirely new traffic is created that follows the traffic pattern from a previously collected Skype network trace. Packets contain simulated headers that match realistic Skype headers. The system also uses a similar approach with HTTP by generating fake HTTP requests from clients, and fake HTTP responses from the server to transmit data (which appear as normal HTTP browsing). The HTTP requests are replays based upon previously collected traces, with header information replaced with the data to be transmitted. The same approach is used for the responses from the server, except the data is hidden within the returned content (such as PDF and JavaScript files).

A third, completely different system is CensorSpoofer [130]. This system, designed for obfuscated web browsing, decouples the upstream and downstream channels. HTTP requests are sent to the server over a low capacity channel such as email or instant messaging. The server responds to the client by mimicking UDP-based VoIP traffic, by mimicking the traffic from a dummy P2P host, more specifically SIP-based VoIP.

All three of these approaches were deemed broken by Houmansadr et al. [53], who proved that all three systems are detectable due to their lack of complete protocol emulation. All three systems do not fully emulate all aspects (for example, error handling) of the protocol that they are attempting to hide as, allowing for detection by comparing the system traffic to legitimate traffic. While this has debunked these three systems, there is ongoing work to make systems like these less detectable. Even though this approach has not been seen in malware yet, it is fully expected that malware will start to take this approach in the near future. At the simplest level, malware that makes use of social networks is starting to adopt this behaviour by mimicking HTTP traffic. A 2013 report from Symantec [121] details an targeted attack against a major internet hosting provider in which malware was installed on linux servers which opened a backdoor. The backdoor operated as a network monitor which scanned all traffic entering the system over SSH (and other protocols). The monitor looked for a certain sequence of characters, namely ":!;". If this flag was seen, the malware extracted encrypted and encoded data which followed. The data could be embedded in any incoming traffic, making it very difficult to detect.

## Future: Namecoin

Another further development that is beginning to be seen in the wild is the use of the Namecoin service. Namecoin is related to Bitcoin, and provides a decentralised method to register and control domain names. Domains that belong to the Namecoin service use the ".bit" top-level domain. The advantage to a malicious user is that is provides the means to anonymously purchase a domain outside the control of any international body. McCardle et al [75] have found malware that is using this service in the wild, and it is expected that ti will become more widespread.

## Future: Esoteric C&C Channels

It is also expected that attacker will make use of further unusual channels for command and control in order to evade controls. A common control is to provide an "air gap" around a machine – i.e. the machine is physically disconnected from any other machine, including the internet. In the perfect situation, this would be a laptop disconnected from the power supply (data can be transmitted through power cabling, a method that is used in the consumer "Powerline" network adapters). Recently, however, Hanspach and Goetz [49] have proposed a design for malware that can operate even in the face of an air gap. The proposed channel is to make use of the microphones and speakers found in most laptops in order to transmit data between machines using inaudible frequencies. Using this channel, a data rate of approximately 20bit/s up to a range of 19.7m can be achieved. By extending the system into a mesh network, multi-hop communication can be achieved. While 20bits/s seems low at first, it is more than enough to transmit small amounts of data such as passwords, banking details or memory dumps.

## D.5 Future Trends

As malware writers attempt to make their malware more resilient to take-down attempts and detection, there are a number of trends that we can expect in the near future.

First, the use of decentralised malware will increase. This will be both down to the additional redundancy provided by a decentralised network, and also the scalability provided by such systems, as it is also expected that botnets will continue to increase in size. We expect that the malware designers will start to use resilient network designs offered by scientific literature.

Second, the use of anonymity services will also increase. As it is harder to avoid detection, and it is getting easier for authorities to locate malware operators, the operators will increasingly want to minimise the risk that they are identified. Services such as Tor will therefore become more widespread.

Finally, although in the wild it is currently extremely rare to find examples, there is a high probability that techniques involving steganography will become more widespread. This will allow the malware to use legitimate services to transmit information, i.e. by hiding in plain sight. This will vastly reduce the effectiveness of most current detection methods.

# C&C Detection

Given the range of C2 design techniques, there is much interest in the design of techniques to localise C2 communication traffic by exploiting its special nature. Detection techniques can be used to carry out such analysis effectively on large scale networks to engage with malicious network activity. In recent years, a number of new techniques have been proposed to mine complex traffic data in order to support correlation and fusion using innovations from the fields of machine learning, semantic analysis, information theory, and traffic analysis. C&C detection falls broadly into two categories: signature-based and non-signature based. In signature-based detection, the detection algorithms are designed to look for known patterns of behaviour collected from malware samples (or "signatures"). These algorithms are often good at detecting the C&C of particular malware, but not so good at detecting new malware. Host-based anti virus systems usually fall into this category. Non-signature based algorithms instead look for anomalies compared to the norm. They are often much more adaptable to new variants of malware, but may not perform as well against known malware. Further to this, there are three different targets for detection, each requiring differing approaches for detection. These are infected hosts, command servers and the communication protocol.

There are two primary measures of the success of a C&C detection, true positive rate(TP) and false positive rate (FP). The true positive rate measures the percentage of malicious samples that are labelled correctly as malware, while the false positive rate measures the number of legitimate samples that are incorrectly labelled as malware.

## E.1 Measurement and Data Collection

When detecting malware C&C, the selection of which data to collect and analyse is extremely important. For example, varying detection methods require different levels of detail in the data. As networks scale, it will get increasingly harder to store all traffic — a requirement of most enterprise C2 detection techniques. Thus, if C2 traffic traces go unrecorded, then detection systems cannot work.

Current measurement techniques have addressed scalability limitations of data collection by developing measurement architectures for aggregation and sampling. However they do so without addressing evasion resilience requirements. Also, little attention has been paid to measurement control mechanisms — tuning measurement in response to C2 evasion.

## E.2 Scalable measurement

Traffic monitoring is performed by routers, commonly using Netflow [17] feature or the sFlow feature. Alternatively, standalone measurement devices [25] observing traffic via network mirroring devices or splitters (optical or electrical) are more flexible than in-router methods. In both cases, traffic traces are exported to collectors which store the traces. Enterprise networks carrying a few tens of terabytes a day, resulting in tens of giga- bytes of flow records are currently manageable as all records can be collected. However, the growth in network speeds might change this in the future. Additionally, C2 designers can attack the measurement system to evade detection. For instance, by flooding the finite-storage data collectors. Network defenders would thus be forced to switch to sampling network traffic as storage of complete traffic flow records could be impossible under conditions of flooding or network congestion.

In the case of ISPs, the volume of traffic flow records is immense. A tier-1 ISP carries several tens petabytes of user traffic per day [1], resulting in hundreds of terabytes of flow records. Even with low storage and transmission costs, storing entire traffic traces beyond a few days is not feasible for ISP traffic while storing the entire traffic including packet data is outright impossible.

The volume of traffic on ISP networks presents a challenge which requires collectors to summarise trace traffic. This can be done either via summarisation techniques or via sampling techniques. Unlike high-level summaries produced by summarisation techniques, sampling techniques produce fine-grained traces that are representative of complete network traffic data. Sampling techniques can support the creation of arbitrary sub-aggregates to support detection techniques that need not be specified at the time that sampling takes place.

The challenge is to achieve the following requirements: 1. Fairness: yield accurate estimates about traffic based on the samples 2. Confirm to the sampling budget – the maximum number of samples to be gathered from data arriving within a specified time period. 3. Timeliness: provide samples in a timely manner to detection mechanisms. Fairness is an important criteria. If the fairness guarantees are weak or non-existent, then the adversary can exploit weaknesses in the sampling algorithm. This can result in the C2 traffic evading the monitoring system, as a consequence detection would fail.

In the rest of this section section, we will briefly outline the main methods of data collection that are used by the detection methods discussed later.

### NetFlow

NetFlow is a network protocol for collecting IP traffic information. Developed by Cisco, NetFlow is used to collect and monitor network traffic flows at the router level. It is the current industry standard for traffic monitoring due to its low overhead but high level of detail.

NetFlow data represents "flows" of traffic. A flow is defined by Cisco to represent a unidirectional sequence of packets with a single source-destination pair. As an example, NetFlow data could consist of the following:

- Source IP
- Destination IP
- Source Port
- Destination Port
- Protocol (e.g. TCP, UDP)

There are numerous other traffic flow features that can also be stored, including timestamps, byte count, and headers. One of the key points, however, is that the actual payload data is not stored. This is due to the fact that the storage requirements would increase dramatically if all data is stored as well (if you imagine a 1Gbps router logging for just 1 day would create 7.2Tb of data to

# C&C Detection (E)

store and process!).

## Honeynets/Malware Traps

Honeynets and malware traps are essentially bait and traps for malware in the wild. A honeynet is typically made up of a number of honeypot nodes, which are machines that run vulnerable (un-patched) software with a goal of becoming infected with malware. The infected machines can then be used to profile malware through wither automatic or human means. This data is one if the primary sources of signatures for signature-based detection methods. Honeynet nodes do not have to be a single machine. It is possible, through the use of virtual machines, to run large volumes of honeynet nodes on a relatively small amount of hardware. It is important to note that often the malware will be prevented from performing illegal activities (DDoS attacks etc) while under the researchers control.

Honeypot techniques have been widely used by researchers. Cooke et al. [22] conducted several studies of botnet propagation and dynamics using Honeypots; Barford and Yegneswaran [9] collected bot samples and carried out a detailed study on the source code of several families; finally, Freiling et al. [38] and Rajab et al. [99] carried out measurement studies using Honeypots. Collins et al. [21] present a novel botnet detection approach based on the tendency of unclean networks to contain compromised hosts for extended periods of time and hence acting as a natural Honeypot for various botnets. However Honeypot-based approaches are limited by their ability to attract botnets that depend on human action for an infection to take place, an increasingly popular aspect of the attack vector [80].

## Sandboxes

A slight variant on a honeynet is a malware sandbox. In this instance, malware is directly installed on a machine and the activities analysed. The main difference with a honeynet, however, is that the owner will also interact with the malware (for example, by mimicking command and control servers). This allows the researcher to gain a much bigger picture of the malware's operation under different situations.

## Reverse Engineering

Perhaps the most labour intensive, reverse engineering is probably the most useful tool in learning about the command and control systems of malware. Many of the examples of command and control systems discussed in the previous section were discovered through reverse engineering. To reverse engineer malware, the researcher will analyse the actual malware binary, and attempt to recover the source code. This can give valuable insights into the operation of the malware, and can even give vital information such as hardcoded C&C server addresses and encryption keys. The main issue is that it can take a very long time to completely reverse engineer a piece of malware (and in some cases it may not be possible at all), and it is a process that is extremely difficult to automate.

## E.3 Signature Based Methods

In signature-based detection methods, malware C&C is detected by looking for known patterns of behaviour, or "Signatures". Signatures are generated for known malware samples, and then new traffic is compared to these signatures. If the new traffic matches a signature, then the traffic is classed as C2 traffic.

Signatures are generated by analysing confirmed C2 traffic collected from various sources. The main sources are honeynets and sandboxes. Malware is run in controlled conditions, and its activity logged. What is logged depends on the detection algorithm being used, but almost every aspect of the malware's behaviour can be included in a signature. Some systems, for example, solely base signatures upon the payload data of packets, while others can cover entire flows and the timings of packets. It is also not the case that one piece of malware will be represented by a single signature, and vice versa. It is often the case that a single malware sample will generate multiple signatures as the conditions on a host machine can vary, which will affect the command and control activity. Conversely, a signature may represent multiple pieces of malware that exhibit very similar behaviour.

## Communication Detection

As we have seen, many malware variants have very particular protocols when it comes to communication. These are often noticeably different to legitimate traffic, both in packet contents and in the behaviour of the communications. This makes signature based detection methods very good for detecting known variants of malware. Many different pieces of malware may also be based upon a common component, meaning that a single signature can be used to detect multiple pieces of similar malware. One possibility for this kind of detection is to produce signatures based upon the contents of packets. It is often the case that packets of data involved in the C&C of malware will be almost identical across multiple hosts. Even though some malware familes use encryption in their communications, that encryption is usually a simple, lightweight algorithm (as the encryption is often for obscurity rather than security), so their are similarities among different ciphertexts. For example, in the work of Rieck et al [103], in which n-gram based signatures are generated for the payloads of malware that is run under controlled conditions in a sandbox. Signatures are also generated for legitimate traffic, and with this method the system can achieve detection rates of close to 95%, with a false positive rate of close to zero when running on a network gateway. Encryption can make the detection of malware traffic much more difficult, especially if the system uses widespread protocols such as HTTP. One approach is then to attempt to decrypt all packets and then perform signature detection on the decrypted contents, as is done by Rossow et al [104]. They take advantage of the fact that in many cases the encryption used is very simple, and often the key for encryption is hardcoded into the malware binary. They keys are fetched by reverse engineering, and then the payloads can be decrypted,

# C&C Detection

ans signature-based detection applied. The obvious down- side to this method is that it requires the labour intensive reverse engineeing step.

Further to this, Rafique et al. [102] proposed a system for large-scale automatic signature generation. The system uses network traces collected from sandboxes and produces signatures for groups of similar malware, covering numerous protocols. This system is able to identify numerous malware example with a high rate, and experiences a low false positive rate due to the specificness of the signatures generated. The signatures are designed to be exported to intrusion detection systems such as Snort for on-line detection.

## Spam Detection

There have also been attempts at performing spam detection based upon the method that the spam email was sent, which is quite often through malware. The work of Stringhini et al [118] utilises the fact that many different mail clients, including malware, introduce slight variations into the standard SMTP protocol. They use this to produce "dialects", which are signatures for each mail client that can represent these variations. Dialects are collected for known sources of spam, including malware, and also for legitimate mail services. It is then a simple case of matching incoming emails to a dialect to make the decision of if the email is spam. In a further piece of work from the same authors [119], they propose a different approach ion BotMagnifier. This system first clusters spam messages according to their content, and then measures the source and destination IP addresses to match clusters to known botnets. This allows for both the enumeration of known botnets, and the discovery of new ones. It is of course the case that many spam campaigns could originate from the same botnet, so clusters that share source IPs are liked to the same botnet. It also is observed that a particular botnet will often target a particular set of destinations, such as one particular country, which is used to add precision.

## Server Detection

Nelms et al. [86] propose ExecScent, a system for identifying malicious domains within network traffic. The system works by creating network traces from known malware samples to create signatures, that can then be compared with network traffic. The sig- natures are not just based upon the domain names, but also the full HTTP requests associated with them. How this system is unique, however, is that the signatures are tailored to the network that they will be used on based upon the background network traffic. This step is extremely useful at reducing the level of false positives by exploiting the fact that different networks will exhibit different browsing behaviour (for example a car manufacturer is unlikely to visit the same websites as a hospital).

## E.4 Non-Signature Based Methods

The main disadvantage of using a signature based detection method is that these detection systems are usually not very effective at detecting new, or updated, malware. Every time a new piece of malware is discovered, or an exiting piece updates itself, the signatures have to be recreated. If the new variant is not discovered, then it is unlikely to be detected by these systems. This is where non-signature based detection comes in. In these systems, the algorithms look for behaviour that is not expected, rather than looking for particular known behaviour, or looking for a specific type of behaviour without the use of signatures.

## Server Detection: DNS

There has been a large amount of work that attempts to provide a detection mechanism that can identify domains associated with malware at the DNS level. As we have seen, DNS is used by a large amount of malware that makes use of a centralised command and control structure.

One proposed detection method is to make use of the reputation of domain names to decide if they are related to malicious activities [6]. In this system (Notos), domains are clustered in two ways. First, they are clustered according to the IP addresses associated with them. Secondly, they are clustered according to similarities in the syntactic structure of the domain names themselves. These clusters are then classified as malicious or not based upon a collection of whitelists and blacklists: domains in a cluster that contains blacklist domains are likely to be malicious themselves. This system is run on local DNS servers and can achieve a true positive rate of 96% and an low false positive rate. In a further piece of work from the same authors as Notos, the idea is vastly expanded to use the global view of the upper DNS hierarchy. In this new system (Kopis) [7], a classifier is built that, instead of looking at the domains' IP and name, looks at the hosts that make the DNS requests. They leverage the fact that malware-related domains are likely to have an inconsistent, varied pool of requesting hosts, compared to a legitimate domain which will be much more consistent. They also look at the locations of the requesters: requesters inside large networks are given higher weighting as a large network is more likely to contain infected machines. When tested, this system was actually able to identify a new botnet based in China, which was later removed from the internet.

DNS is also used in another way by malware controllers that we have not yet mentioned. One feature of DNS is DNS blacklists (DNSBL). These are used by spam filters to block emails from known malicious IPS. The malware controllers will often query these blacklists for IPs under their control to test their own networks [101]. The behaviour of a botmaster performing DNS lookups for his own hosts will differ from legitimate use of DNSBLs. For example, a malicious host performing DNS lookups on behalf of the controller will perform lots of queries, but will not be queried itself, while a legitimate service will receive incoming queries. This behaviour is relatively easy to detect by simply looking for queries that exhibit this behaviour.

Paxson et al [89] attempt to provide a detection mechanism that leverages the amount of information transmitted

# C&C Detection

over a DNS channel in order to detect suspicious flows. The system allows for a upper bound to be set, any DNS flow that exceeds this barrier is flagged for inspection. The upper bound can be circumvented by limiting flows, but this has an impact on the amount of data exfiltration/command issuing that can occur. The system looks primarily at data included within domain names, but also looks at interquery timings and DNS packet field values, both of which can provide low capacity channels.Several other works seek to exploit DNS usage patterns. Dagon et al. [26] studied the propagation rates of malware released at different times by redirecting DNS traffic for bot domain names. Their use of DNS sinkholes is useful in measuring new deployments of a known botnet. However, this approach requires a priori knowledge of botnet domain names and negotiations with DNS operators and hence does not target scaling to networks where a botnet can simply change domain names, have a large pool of C&C IP addresses and change the domain name generation algorithm by remotely patching the bot. DNS blacklists and phishing blacklists [110], while initially effective have are becoming increasingly ineffective [100] owing to the agility of the attackers. Much more recently, Villamar et al. [128] applied Bayesian methods to isolate centralised botnets that use fast-flux to counter DNS blacklists, based on the similarity of their DNS traffic with a given corpus of known DNS botnet traces.

## Fast Flux

As we recall, in a fast flux network the command and control server is hidden behind a proxy of numerous compromised hosts. Performing DNS queries on the domain of the server will return a large, and constantly changing, set of IP addresses. As you may expect, this type of behaviour is relatively easy to detect.

As we discussed, there are some differences between fast-flux service networks (FFSNs) and content delivery networks (CDNs) [51]. To detect a FFSN is a simple process, due to the two characteristics of an FFSN: short TTL values in DNS responses and non-overlapping DNS responses. If DNS traffic is monitored, then by simply looking for DNS responses for domains that meet this criteria will indicate a possible FFSN. This can also be done manually for individual suspect domains by generating multiple DNS queries. This will give two pieces of information. The main result is that domains can be identified as being behind FFSNs and therefore added to blacklists. Secondly, the returned IP addresses will be those of likely compromised machines, which are quite possibly part of a botnet. This list can be compared with internal networks to identify and mitigate compromised machines, and also enumerate the botnet.

It is also possible to automatically detect which domains belong to the same FFSN. The work or Perdisci et al [90] applies clustering to domains so they are grouped according to overlap in the returned IP addresses. By then comparing the clusters to previously labelled data, they can then be classified as flux or non-flux, revealing domains that make use of the same network.

## Host Detection

An interesting system for host detection is BotHunter [44]. BotHunter is a system for identifying compromised hosts based upon the actions they perform, more specifically the pattern of infection and initial connection to a command and control server. There are 5 steps to this patter: inbound scan, inbound exploit, binary download, outbound C&C communication and outbound infection scanning (for propagation). These steps are identified as being an good generalisation of the typical infection model for a botnet (although some botnets will obviously leave out or add extra steps). The system works by correlating IDS alerts and the payloads of request packets. These are used to identify hosts performing the 5 steps, and if a host is found to perform certain combinations of these within a time period, they are identified as a possible bot. The timer is used as legitimate services may give the appearance of performing one of these steps. There are two conditions for a host to be labelled as compromised. The first is that it has been the victim of an inbound exploit, and has at least one occurrence of outward C&C communication or propagation. The second is that it has at least two distinct signs of outward bot coordination or attack propagation. This system can achieve 95% detection rates, and low false positive rates. The downside, however, is that as it is heavily reliant on detecting the behaviour of existing botnets it can be evaded by slowing down the infection process to fall outside the time limits. BotHunter is available as an open source product. The BotHunter authors produced a further system, BotMiner [43], that detects infected hosts without previous knowledge of botnets. In this system, bots are identified by clustering hosts that exhibit similar communication and (possible) malicious activities. The clustering allows hosts to be groups according to the botnet that they belong to as hosts within the same botnet will have similar communication patterns, and will usually perfrom the same activities at the same time (such as a DDoS attack).

Finally, there are also schemes that combine network and host-based approaches. The work of Stinson et al. [112] attempts to discriminate between locally-initiated versus remotely-initiated actions by tracking data arriving over the network being used as system call arguments using taint tracking methods. Following a similar approach, Gummadi et al. [48] whitelist application traffic by identifying and attesting humangenerated traffic from a host which allows an application server to selectively respond to service requests. Finally, John et al. [61] present a technique to defend against spam botnets by automating the generation of spam feeds by directing an incoming spam feed into a Honeynet, then downloading bots spreading through those messages and then using the outbound spam generated to create a better feed.

## Graph-based approaches

Several works [20, 56, 57, 60, 138] have previously applied graph analysis to detect botnets. The technique of Collins and Reiter [20] detects anomalies induced in a graph of protocol specific flows by a botnet control traffic. They suggest that a botnet can be detected

# C&C Detection

based on the observation that an attacker will increase the number of connected graph components due to a sudden growth of edges between unlikely neighbouring nodes. While it depends on being able to accurately model valid network growth, this is a powerful approach because it avoids depending on protocol semantics or packet statistics. However this work only makes minimal use of spatial relationship information. Additionally, the need for historical record keeping makes it challenging in scenarios where the victim network is already infected when it seeks help and hasn't stored past traffic data, while our scheme can be used to detect pre-existing botnets as well. Illiofotou et al. [56,57] also exploit dynamicity of traffic graphs to classify network flows in order to detect P2P networks. It uses static (spatial) and dynamic (temporal) metrics centred on node and edge level metrics in addition to the largest-connected-component-size as a graph level metric. Our scheme however starts from first principles (searching for expanders) and uses the full extent of spatial relationships to discover P2P graphs including the joint degree distribution and the joint-joint degree distribution and so on.

Of the many botnet detection and mitigation techniques mentioned above, most are rather *ad hoc* and only apply to specific scenarios of centralised botnets such as IRC/HTTP/FTP botnets, although studies [42] indicate that the centralised model is giving way to the P2P model. Of the techniques that do address P2P botnets, detection is again dependent on specifics regarding control traffic ports, network behaviour of certain types of botnets, reverse engineering botnet protocols and so on, which limits the applicability of these techniques. Generic schemes such as

BotMiner [43] and TAMD [135] using behaviour based clustering are better off but need access to extensive flow information which can have legal and privacy implications. It is also important to think about possible defences that botmasters can apply, the cost of these defences and how they might affect the efficiency of detection. Shear and Nicol [87, 107] describe schemes to mask the statistical characteristics of real traffic by embedding it in synthetic, encrypted, cover traffic. The adoption of such schemes will only require minimal alterations to existing botnet architectures but can effectively defend against detection schemes that depend on packet level statistics including BotMiner and TAMD.

## E.5 Host Detection

An initial defence against botnets is to prevent systems from being infected in the first place. Anti-virus software, firewalls, filesystem intrusion detection systems, and vulnerability patches help, but completely preventing infection is very difficult task. Malware authors use encryption [136] and polymorphism [123] among other obfuscation techniques [123] to thwart static analysis based approaches used by anti-virus software. In response, dynamic analysis (see Vasudevan et al. [126] and references therein) overcomes obfuscations that prevent static analysis. Malware authors have countered this by employing trigger based behaviour such as bot command inputs and logic bombs which exploit analyzer limitations of only observing a single execution path. These limitations are overcome by analyzing multiple execution paths [14, 78], but bots may in turn counter this using schemes relying on the principles of secure triggers [39, 109]. In order to remain invisible to detection, bots can also use a variety of VM (Virtual Machine) based techniques for extra stealth, such as installing virtual machines underneath the existing operating system [65] to prevent access from software running on the target system and being able to identify a virtual analysis environment including VMs and Honeypots [36]. Graph analysis techniques have also been used in host-based approaches. BLINC [62] is a traffic-classification method that uses "Graphlets" to model flow characteristics of a host and touches on the benefit of analyzing the "IP social-network" of a machine. Graph analysis has also been applied to automated malware classification based on function call graphs [54].

One of the areas that is most important to organisations is to identify hosts that are infected malware so appropriate actions can be taken. It is important to note here that we are only interested in host detection through the command and control actions of the malware, NOT the actual infection of the malware itself through binary detection (as is covered by anti-virus software).

# Controls for C&C

Over the years, a number of security standards, recommendations, and best practices have been proposed to address security risks. In particular, the Council on CyberSecurity (CCS) publishes and manages the "Critical Controls for Effective Cyber Defence v4.1" [23], a list of key actions that organisations should take to detect, block, or mitigate attacks. The controls are informed from experience with actual attacks, as provided by a broad range of contributors to the list, and are designed so that they can be implemented, enforced and monitored largely in an automated fashion. These controls are recommended by UK Government for improving cyber defences in all organisations.

Hereinafter, we will review the Controls in the context of detecting and disrupting C2 activity. We will base our review on version 4.1 of the Controls, the latest available at the time of writing. More precisely, we will highlight the controls that appear suited at defending against C2: we will reflect on their effectiveness and on their practical applicability on the basis of the C2 techniques that we have discussed so far.

## F.1 Controls for C2 Detection

### Critical Control 5: Malware Defences

Control 5 is a very broad control that encompasses processes and tools for detecting, preventing, or correcting the installation and execution of malicious software on all devices of an organisation.

Some of the actions it recommends are related to the prevention of infections (e.g., keeping systems and defence tools up to date, disabling auto-run mechanisms and preforming automatic scans of removable media, emails, and web pages, deploying anti-exploitation techniques). Several actions can instead be used to specifically detect and disrupt C2 activity:

- Monitoring all inbound and outbound traffic on a continuous basis. The control specifically suggests to watch large transfers of data or unauthorised traffic, which may happen during the exfiltration phase of an attack.

- Detecting anomalies in network flows. The control recommends to look for anomalies in the network traffic which may be indicative of malware activity (such as C2 communications) or of compromised machines.
- Logging DNS queries and applying reputation checks. The control suggests to monitor DNS requests for attempts to resolve known malicious domains or attempts to contact domains with poor reputation.

### Critical Control 13: Boundary Defence

Control 13 is concerned with detecting and preventing information flows at an organisation boundaries that may violate the organisation's security policies. More specifically, it can be used to identify signs of attacks and evidence of compromise.

The practical actions that this control recommends include:

- Using blacklists to deny communication from internal machines toward known malicious hosts.
- Storing network traffic and alerts in logs analytics systems for further analysis and inspection.
- Deploying NIDS to monitor the network traffic looking for signs of infection.
- Capturing and analysing netflow data to identify anomalous activity.
- Configuring the network so that all outgoing traffic passes through a "choke point" and so that it can be segmented to prevent and contain infections.

### Critical Control 17: Data Loss Prevention

The goal of control 17 is to track, control, prevent, and correct data transmissions and storage that violate an organisation's security policy. Since stealing sensitive data is the final objective of most targeted attacks, the recommendations of this control are clearly relevant in the context of C2 activity.

A number of actions described as part of this control can be effectively used to detect and mitigate C2 channels:

- Deploying data loss prevention (DLP) tools at the perimeter, to identify sensitive data leaving the organisation premises. These tools often search the traffic for keywords or data formats that are associated with sensitive data.
- Detecting the unauthorised use of encryption in network traffic. The rationale here is that malware may use encryption to exfiltrate sensitive data bypassing tools (such as DLPs) that rely on the inspection of traffic content.
- Blocking access to known file transfer and email exfiltration sites.
- Searching for anomalies in traffic patterns.

## F.2 Controls for C2 Disruption

### Critical Control 19: Secure Network Engineering

Control 19 prescribes a set of actions to broadly create an infrastructure that can withstand attacks. In particular, the following actions are relevant to the task of disrupting C2 activity:

- Segmenting the network according to trust zones. This activity can be particular beneficial if it possible to clearly separate high-risk components of the network (e.g., parts that are particularly exposed to attacks) from high-value components (e.g., those that store sensitive data).
- Designing an infrastructure that allows the rapid deployment of new access controls, rules, signatures, etc. This is especially important to reap the benefits of other controls we have discussed: for example, to deploy new blacklists that have been available or to update the signatures of indicators of compromise used in network-based monitors.
- Ensure that clients query internal DNS servers, which can be monitored and whose replies can be manipulated to, for example, prevent access to known malicious or unauthorised domains.

# Controls for C&C

## F.3 Other Controls

The Critical Control list includes a few other controls that are not immediately related to the detection or disruption of C2 activity, but that are often associated to the defence against targeted attacks. More precisely, *Control 9 (Security Skills Assessment and Appropriate Training to Fill Gaps)* recommends training employees and organisation members to be aware of attacks. Intuitively better awareness can help avoiding human mistakes. However, the effectiveness of security training in general is debated [108], and the characteristics of targeted attacks may make training even less effective (e.g., attacks are more likely to resemble normal activity). Some case studies describing training programs specifically designed with targeted attacks in mind have been described in the literature [111].

*Critical Control 18 (Incident Response and Management)* indicates a list of actions for responding to incidents. Clearly, having a well defined plan to deal with the detection of C2 activity is necessary to avoid or minimise the damages of an attack or ongoing infection.

Finally, *Critical Control 20 (Penetration Tests and Red Team Exercises)* should also be taken in account in the context of C2 activity as a way to test the effectiveness of the techniques and tools used within an organisation. In particular, such security exercises should test whether attempts to set up C2 channels, using both known and new techniques or variations on existing techniques, would be detected by the other controls employed by the organisation.

## F.4 Discussion

### Limitations

Our review of the Critical Controls shows that while they do include sensible advice on defending against C2 activity, they also have some limitations that may hinder their effective adoption. For the most part, these limitations seem a consequence of the general nature of the 20 Critical Controls, which are not tailored to C2 activity specifically. First, controls are often extremely broad, encompassing a wide variety of technologies and approaches. For example, the activities listed in Control 5 encompass whole sectors of the information security industry, ranging from anti-virus technologies, intrusion detection systems, reputation systems, and anomaly detection. This is not a problem per se: the use of orthogonal mechanisms ("defence in depth") has long been considered good practice. However, extracting techniques that are specific for C2 detection and disruption among the full list of controls may become daunting.

Similarly, activities that are relevant for C2 detection are scattered through several controls, which makes it more difficult for someone focusing on C2 to ensure that all relevant controls have been implemented or considered. Finally, the Controls document provides little discussion of the limitations inherent in the controls it proposes. While the metric and test sections in each control provide a discussion of how to measure and test the effectiveness of a control, it may be easy for a reader to focus on the defensive mechanisms rather than on the results that they provide.

### Generalization of controls for C2 detection and disruption

From our discussion of C2 techniques and defences, it is evident that most approaches to the detection of C2 activity rely on monitoring network traffic and applying some form of detection algorithm on it. The Security Controls do include activities that lead organisations toward this approach to security; here, we will generalise and comment on these recommendations:

- Monitor all inbound and outbound traffic. More precisely, it is important to inspect inbound traffic for signs of attacks that may lead to an infection, for ex- ample, drive-by-download or spear phishing attacks. Outbound traffic should be analysed looking for indications that a C2 channel has been established (data ex- filtration, Command & Control check-in, etc.)
- Monitor network activity to identify connection attempts to known-bad end points, i.e., IPs and domains that are known to be used in attacks. The rationale is that access to these endpoints can be prevented, assuming that appropriate mechanisms are in place (e.g., firewalls). The key aspect here is of course that of creating and maintaining up-to-date lists of malicious endpoints. Different approaches to create and evaluate such lists have been proposed both in the academia and in the commercial sector [32, 58, 68, 95, 113, 124].
- Identify and inspect anomalies in the network traffic. The rationale is that targeted attacks rely on infrastructure that is less likely to be included in generally- available lists of malicious endpoints or to use C2 techniques (e.g., protocols) that are used also by general malware. Then, focusing on detecting anomalous traffic would enable defenders to catch these novel threats. There are two assumptions underlying this recommendation: targeted attacks result in anomalous traffic and anomalous traffic is an indication of compromise. Both assumptions may need to be re-evaluated from time to time: we have seen that attackers are devising new methods to "blend in" with the normal traffic; the characteristics of traffic on a network may change as new services and devices are introduced.
- Collect specific subsets of network traffic, in particular DNS queries and netflow data. A motivation for this recommendation is that it may be easier to collect such data, rather than setting up a full network monitoring system. As we have seen from our literature review, several approaches have been devised to identify C2 traffic based on these inputs.
- Architect the network in such a way that simplifies traffic monitoring and the activation of responses to attacks. For example, by having a single choke point where all traffic passes through, an organisation can simplify the full collection of traffic and its inspection. As another example, network segmentation can help keeping separated networks of different trust values (e.g., networks hosting front- facing

# Controls for C&C

servers vs. those hosting internal services). In addition, the use of rate limiting techniques may slow down attackers as they try to exfiltrate data and increase the window of time in which a detection can occur.

**Risks**

There are several factors that may limit the effectiveness of a control. Attackers are always looking for ways to "remain under the radar" and avoid detection. For example, to limit the effectiveness of content analysis techniques, they may use encrypted communication protocols, or they may adapt their C2 traffic so that it resembles regular traffic seen on a network. To thwart controls that call for matching traffic (e.g., connection endpoints, DNS queries) against lists of known malicious entities, attackers refrain from re-using artefacts (such as actual attack vectors, servers, and domain names) in multiple attacks. To work around anomaly detection approaches, attackers may make their activities, in particular their C2 traffic, similar to benign traffic.

**Practical matters**

We will conclude our review of security controls with a discussion of some non technical issues that may face an adopter of the controls. For example, an organisation may not have sufficient resources (staff, time, or money) to apply a control in its entirety. In addition, implementing a control may require changes to or collaboration from a multitude of departments or groups inside organisation. For example, monitoring DNS queries may require that the security group interacts with the networking group. It would be helpful to have some guidance on addressing such issues, perhaps in the form of case studies.

An approach that we have seen applied successfully to the introduction of new controls for C2 activity could be summarised as "start small, measure, and scale up". An organisation does not need to apply a control throughout its entire infrastructure (*start small*): for example, it could choose to initially protect a subset of users, such as a high-risk group, or a group that is tolerant to initial experimentation with potentially higher than normal false positive rates. Similarly, an organisation could choose to focus on a specific type of traffic (e.g., DNS) that has smaller performance requirements and still a good potential of leading to the detection of C2 channel activity. After the initial, limited implementation of a control, its effectiveness should be assessed (*measure*). If successful, the control could be extended to larger portions of the organisation (*scale up*).